\documentclass[superscriptaddress,aps,pra,twocolumn]{revtex4}

\usepackage{graphicx}% Include figure files
\usepackage{dcolumn}% Align table columns on decimal point
\usepackage{bm}% bold math
\usepackage{bm}
\usepackage{ulem}
\usepackage{epsfig}
\usepackage{graphicx}
\usepackage{amssymb,amsmath,amsbsy,amsgen,amsfonts}    
\usepackage{dcolumn}
\usepackage{amsthm}
\usepackage{mathrsfs}
\usepackage{latexsym}
\usepackage{array}
\usepackage{color}
\usepackage{amstext}
\allowdisplaybreaks[1]
\usepackage{txfonts}
\usepackage{pifont}
\usepackage{braket}
\usepackage{epstopdf} %for texshop on mac

\newcommand{\be}{\begin{equation}}
\newcommand{\ee}{\end{equation}}
\newcommand{\ba}{\begin{array}}
\newcommand{\ea}{\end{array}}
\newcommand{\bqa}{\begin{eqnarray}}
\newcommand{\eqa}{\end{eqnarray}}

\DeclareSymbolFont{symbols}{OMS}{cmsy}{m}{n}

\begin{document}

\title[]{Experimental measurement of kinetic parameters using quantum plasmonic sensing}
\author{K. T. Mpofu}
\affiliation{Catalysis and Peptide Research Unit, School of Health Sciences, University of KwaZulu-Natal, Durban 4041, South Africa}

\author{C. Lee}
\affiliation{Quantum Universe Center, Korea Institute for Advanced Study, Seoul 02455, Republic of Korea}
\affiliation{Korea Research Institute of Standards and Science, Daejeon 34113, Republic of Korea}

\author{G. E. M. Maguire}
\affiliation{Catalysis and Peptide Research Unit, School of Health Sciences, University of KwaZulu-Natal, Durban 4041, South Africa}
\affiliation{School of Chemistry and Physics, University of KwaZulu-Natal, Durban 4041, South Africa}

\author{H. G. Kruger}
\affiliation{Catalysis and Peptide Research Unit, School of Health Sciences, University of KwaZulu-Natal, Durban 4041, South Africa}

\author{M. S. Tame}
\email{markstame@gmail.com}
\affiliation{Laser Research Institute, Department of Physics, Stellenbosch University, Private Bag X1, Matieland 7602, South Africa}

\date{\today}

\begin{abstract}
Kinetic models are essential for describing how molecules interact in a variety of biochemical processes. The estimation of a model's kinetic parameters by experiment enables researchers to understand how pathogens, such as viruses, interact with other entities like antibodies and trial drugs. In this work, we report a simple proof-of-principle experiment that uses quantum sensing techniques to give a more precise estimation of kinetic parameters than is possible with a classical approach. The interaction we study is that of bovine serum albumin (BSA) binding to gold via an electrostatic mechanism. BSA is an important protein in biochemical research as it can be conjugated with other proteins and peptides to create sensors with a wide range of specificity. We use single photons generated via parametric down-conversion to probe the BSA-gold interaction in a plasmonic resonance sensor. We find that sub-shot-noise level fluctuations in the sensor signal allow us to achieve an improvement in the precision of up to 31.8\% for the values of the kinetic parameters. This enhancement can in principle be further increased in the setup. Our work highlights the potential use of quantum states of light for sensing in biochemical research. 
\end{abstract}

\maketitle

\section{\label{Introduction}Introduction}
The measurement of kinetic parameters plays an important role in characterizing the physical mechanisms underlying molecular interactions, enabling the development of vaccines, drugs and cancer treatments~\cite{Pollard2010}. Plasmonic sensors are optical sensors that are widely used in industry for studying molecular interactions due to their high sensitivity~\cite{Homola1999,Homola1999b,Homola2003,Homola2006,Li2015,Salazar2018,Xiao19}, label-free approach~\cite{Soler2019} and specificity~\cite{Homola2008}. Despite the advantages of using plasmonic sensors for studying biochemical processes, the precision in the measurement performed by these sensors is starting to reach a fundamental limit known as the shot-noise limit~\cite{Homola2009}. This is due to the statistical structure of the light sources used in the sensors. Achieving a better precision is possible by using quantum light sources with reduced noise~\cite{Giovannetti11,Taylor16,Degen17,Pirandola18}. Recent work has explored the use of quantum light in plasmonic sensing, introducing `quantum plasmonic sensors' as a new approach to biosensing~\cite{Lee2021}. Several studies have shown theoretically~\cite{Lee2017,Tame19} and experimentally~\cite{Fan2015,Pooser2016,Dowran2018,Lee2018,Peng2020,Zhao2020} an enhancement in the estimation precision of static parameters using quantum plasmonic sensors. Most recently, theoretical work has shown that the enhancement in precision should carry over to estimating kinetic parameters~\cite{Mpofu2021}. However, so far there has been no experimental confirmation. 

In this work we report a proof-of-principle experiment that demonstrates a quantum enhancement in the precision of estimating kinetic parameters. We use single photons as the quantum light source, which are generated by parametric down-conversion. The single photons in our experiment probe a plasmonic resonance sensor which is set up to monitor the interaction of the protein bovine serum albumin (BSA) with gold. BSA is a protein that is widely used in biochemical studies, as it is capable of binding to many types of antibodies and drugs~\cite{Alsamamra2018}. Thus, it is an informative first test case in the study of whether a quantum enhancement can be achieved in the precision of measuring kinetic parameters. The adsorption of BSA to a gold surface occurs via an electrostatic interaction~\cite{Brewer2005,Beykal2015}, and by using a temporal signal from the single photons transmitted through the plasmonic resonance sensor for different concentrations of BSA, we estimate the association and dissociation kinetic parameters for the interaction. Due to the reduced noise of the single photon statistics we find that an improvement of up to 31.8\% in the precision of the values of the association and dissociation parameters is possible, in line with theoretical predictions. Our work shows that quantum light sources can be used for practical sensing of kinetic parameters with an improved precision compared to a classical approach. This may open up new possibilities for designing quantum-based sensors for high-precision biochemical research.

In Section II we describe our optical setup for quantum plasmonic sensing and outline how the kinetic parameters for the interaction model of BSA binding to gold can be extracted from the sensor signal. In Section III we provide the results of our experimental study, comparing the precision in the estimation of the kinetic parameters with those from an equivalent classical setup. In Section IV we summarize our main findings and provide an outlook on future work.
\begin{figure*}[t]
\centering
\includegraphics[width=18cm]{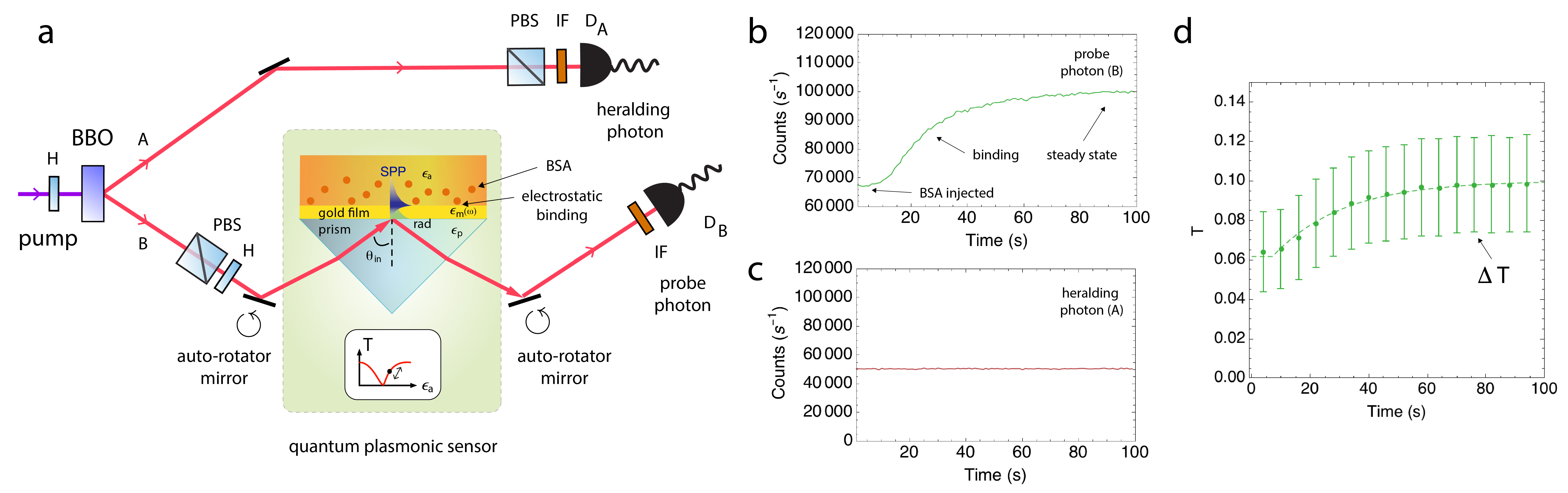}
\caption{Experimental measurement of kinetic parameters using a quantum plasmonic sensor. (a) Setup with quantum light source and plasmonic resonance sensor. Pairs of photons are generated via parametric down-conversion using a nonlinear Beta-Barium-Borate (BBO) crystal. The photon in the top mode A is detected and `heralds' the presence of a single photon in the bottom mode B. This photon is used to probe the transmission response of a plasmonic resonance sensor comprised of a prism with gold film attached by index matching oil. BSA is injected into the sensing region above the gold and the change in refractive index ($n_a=\sqrt{\epsilon_a}$) causes a change in the transmission, $T$, over time as the BSA binds to the gold surface (see inset). {\sf H} is for half-wave plate, {\sf PBS} for polarizing beamsplitter, {\sf IF} for interference filter and ${\sf D_{i}}$ is a single-photon avalanche photodetector in mode ${\sf i}$. (b) Singles counts at the probe detector for the case of 1.5\% BSA concentration injected into the region above the gold surface. The counts vary as the BSA binds, eventually reaching a steady state. (c) Singles counts at the heralding detector. (d) Dependence of the estimated transmission $T$ (points) of the sensor with time and its associated estimation precision $\Delta T$ (error bars) for 1.5\% BSA concentration. The transmission and its precision are obtained based on detection events at detector B given that detector A recorded, or `heralded', a detection event within a 4~ns time window. This ensures that the photons in mode B are single photons with corresponding noise. The dashed line is a nonlinear fit to the points, see next section for details.}
\label{fig1} 
\end{figure*}

%%%%%%%%%%%%%%%%%%%%%%%%%%%%
%%%%%%%%%%%%%%%%%%%%%%%%%%%%
%%%%%%%%%%%%%%%%%%%%%%%%%%%%
%%%%%%%%%%%%%%%%%%%%%%%%%%%%
\section{Experimental scheme}

%%%%%%%%%%%%%%%%%%%%%%%%%%%%
%%%%%%%%%%%%%%%%%%%%%%%%%%%%
\subsection{Experimental setup}

In Fig.~\ref{fig1}~(a) we show the experimental setup used for measuring kinetic parameters with a quantum plasmonic sensor. It consists of a quantum light source and a plasmonic resonance sensor (shown in the shaded green area). For the quantum light source, pairs of photons are produced using spontaneous parametric down-conversion in a nonlinear Beta-Barium-Borate (BBO) crystal~\cite{Burnham70,Hong1986}. A pump photon with wavelength centered at 405~nm from a continuous wave laser at a power of 20mW (Coherent OBIS 405~nm) has its polarization set to vertical using a half-wave plate (H) and is sent into a 3~mm long BBO (from Newlight Photonics). The photon has a non-zero probability to be down-converted by a second-order nonlinear process into a pair of horizontally polarized photons with their wavelength centered at 810~nm. The optical axis of the BBO is cut such that one of the photons exits the crystal at $+3$ degrees from the pump forward direction into mode A and the other photon exits at $-3$ degrees into mode B. The polarizations of photons in both modes are cleaned up using a polarizing beamsplitter (PBS) which transmits horizontally polarized light only, while the spectral bandwidth of the photons is selected using interference filters (IF) ($\lambda_0=810$~nm, $\Delta \lambda =10$~nm) placed before fiber couplers leading to single-photon avalanche photodetectors ${\sf D_{A}}$ and ${\sf D_{B}}$ (Excelitas SPCM-AQRH-13-FC). The detection of a photon in mode A `heralds' the presence of a single photon in mode B. This single photon is used to probe the transmission response of the plasmonic resonance sensor. Finally, an iris is placed in mode A to improve the spatial selection of the down-conversion and give a better correlation with the photons in mode B. This is important for maximizing the overall transmission efficiency of the setup, as described later.

The plasmonic resonance sensor is made up of a prism (BK7 material) and a microscope slide coated with a 50~nm gold film with a titanium adhesion layer (30020011 from Phasis S\`arl). The slide is attached to the prism by index-matching oil (56822-50ml from Sigma Aldrich). The sensor is initially operated under static and not flow conditions -- a buffer solution of deionized water is placed on the gold film using a silicone cavity with dimensions 2~cm$\times$4~cm$\times$1~mm. The gold film supports surface plasmon polaritons (SPPs), which are excitations of light coupled to electron charge oscillations on the top surface of the gold~\cite{Maier07}. At a particular angle of incidence, set by using a mirror on an automated-rotator, the single photons satisfy mode-matching conditions for coupling to single SPPs~\cite{Tame08}. A key mode-matching condition required is that the polarization of the photons is set to be parallel to the plane of incidence. This is achieved by a half-wave plate before the auto-rotator. The photon-SPP coupling, or resonance, corresponds to a dip in the transmission of mode B around the optimal angle, corresponding to a decrease in the number of photons reflected from the gold surface and detected. This occurs as single photons mainly couple to single SPPs which propagate along the gold surface instead. A second auto-rotator is used to guide any reflected photons into the detector as the incident angle is varied so that the transmission can be determined and the location of the optimal resonance angle of the dip can be found.

For sensing, the incidence angle, $\theta_{\rm in}$, is decreased slightly from resonance so that the transmission in mode B increases and it sits on a steep part of the left hand side of the angular dip (near the inflection point). At this point single photons are partially transmitted and partially converted into SPPs. The transmission, $T$, here is most sensitive to changes in the refractive index, $n_a=\sqrt{\epsilon_a}$ (where $\epsilon_a$ is the permittivity), of the medium above the gold film, as shown as a point in the inset of Fig.~\ref{fig1}~(a). In the inset, the inflection point is on the right hand side of the refractive index dip. This is in contrast to the left hand side of the angular dip used, which is due to the incident angle being fixed and the refractive index changing and causing the resonance behavior instead~\cite{Maier07}.

The protein BSA is in a powder form (A2153-10G from Sigma Aldrich) and is mixed with deionized water to give a fixed concentration. The solution is then injected into the silicone cavity above the gold surface using a syringe. As the BSA binds to the gold over time the change in the refractive index above the surface causes a change in the transmission, $T$. This change can be seen in the unheralded `singles' counts detected in mode B for the probe photon (counts in mode B that are not conditioned on a photon in mode A), as shown in Fig.~\ref{fig1}~(b). On the other hand, the transmission in mode A for the heralding photon remains constant, as can be seen in the singles counts for mode A in Fig.~\ref{fig1}~(c). The lower overall count rate compared to mode B is due to an additional iris on mode A that is used to improve the spatial selection of photons in that mode and provide a better correlation with photons in mode B. The time dependent transmission profile, $T(t)$, of mode B is called a sensorgram, and from it we can use a nonlinear fit to extract out kinetic parameters for the binding interaction between the BSA and gold~\cite{Xiao19}. We focus on studying the changes in $T$ due to binding and steady state kinetic behavior. Further details about how $T$ is related to the kinetics and how the extraction procedure is performed are given in the next section.

While the detected singles counts in mode B show a dependence on the temporal transmission of the sensor, the noise in the counts is shot-noise limited due to the probabilistic nature of the down-conversion process at the BBO, which follows a Poissonian process. This stems from the fact that each pump photon that produces a photon pair originates from a continuous wave coherent state of the pump laser, which has a Poissonian distribution of photons~\cite{Loudon00}. In order to remove the noise and go below the shot-noise limit, we use heralded photons in mode B, where a detection of a photon in mode A heralds the presence of a single photon in mode B. The detection in mode B must occur within a coincidence time window for the photon to be considered to come from the same pair~\cite{Hong1986}, which we set as 4~ns in our experiment. In principle, this removes the shot-noise completely, however, the heralding is not perfect due to loss in mode B, which results in some heralded photons not making it to the detector. The resulting noise is binomial~\cite{Gardiner00}, which crucially is smaller than the shot-noise depending on the amount of loss and the transmission due to the plasmonic sensor~\cite{Lee2021}. 

It is important to note that while the $N$-photon number state is known to be optimal for reducing the noise in this transmission scenario~\cite{Nair2018,Tame19,Lee2021}, single-photon states ($N=1$) give the same relative amount of noise reduction as their classical counterpart: the weak coherent state with a mean photon number of 1. Moreover, the use of $N$ single photons can achieve the same limit that would be obtained by the $N$-photon number state~\cite{Lee2021}, as each photon in the state undergoes an independent Bernoulli sampling~\cite{Gardiner00}. This equivalence allows the use of single photons as a practical alternative to $N$-photon number states in quantum sensing. Thus, quantum sensing in this scenario does not need to take advantage of entanglement, nor higher-order Fock states~\cite{Lee2021}.

%%%%%%%%%%%%%%%%%%%%%%%%%%%%
%%%%%%%%%%%%%%%%%%%%%%%%%%%%
\subsection{Statistical data processing}

For a given period of time, we let $\nu$ be the number of single photons sent to the sensor in mode B (probe photons), which are heralded by a detection of a photon in mode A. We then measure the number of transmitted photons in mode B, denoted as ${N}_{\rm t}$. The transmission for that period of time is then ${T = {N}_{\rm t} / \nu}$. Based on the slow time scale on which the sensorgram $T(t)$ changes (see Fig.~\ref{fig1}~(b)), the rate of single-photon counts at detector A (which is $5\times 10^{4}~{\rm s}^{-1}$, see Fig.~\ref{fig1}~(c)), and the rate of pairs of photons detected when the plasmonic sensor in mode B is at the inflection point ($3.3\times 10^{3}~{\rm s}^{-1}$), we use a total collection period of 6 seconds, within which we carry out $\mu=2 \times 10^3$ sets of $\nu =150$ independent and identical samplings, while $T$ remains approximately constant. As $\nu$ is fixed, we require the ability to `time tag' detection events at detectors ${\sf D_{A}}$ and ${\sf D_{B}}$. This allows us to collect $\nu$ counts at ${\sf D_{A}}$ and record the number of times a count occurs at ${\sf D_{B}}$ after a count occurred at ${\sf D_{A}}$ within the coincidence window. For the time tagging, we use a TimeHarp 260 PICO (PicoQuant), which has two independent electronic channels with a 25~ps temporal resolution. The value of $\mu$ was chosen as it produced a steady standard deviation of the mean transmissions obtained from each set of $\nu$ samplings. 

The mean transmission is used as an estimator and for set $i$ it is given by $T_i = {N}_{{\rm t}, i} / \nu$. The expected mean transmission for a 6 second period is then ${ \langle T \rangle =\sum_{i=1}^{\mu} \frac{1}{\mu} }{T_i}$. The precision of the estimation of the mean transmission for a single set of $\nu$ samplings, $\Delta T$, can be found from the standard deviation, or uncertainty, of the mean of the sets, and is given by
\be
\Delta {T}=\sqrt{{ \frac{1}{\mu} \sum_{i=1}^{\mu} ({T_i - \langle T \rangle)^2 }}}.
\label{eq:expDT}
\ee
In Fig.~\ref{fig1}~(d) we show an example sensorgram from our experiment for the case of 1.5\% BSA injected into the cavity region above the gold surface. The points at 6 second intervals represent $\langle T \rangle$ and the error bars represent $\Delta {T}$. The maximum mean transmission through mode B in our setup is approximately 10\% when the SPP angular dip is off resonance to the left of the dip where total internal reflection takes place and the sensor has $T=100$\% ideally. 

The overall 10\% efficiency of our sensor is due to a number of factors, the main one being the prism itself (PS912 Thorlabs). The prism is right-angled and made of N-BK7, with sides of length 40~mm, a base of 56.5~mm and width of 40~mm. The size was chosen in order to support the gold microscope slide. Photons in the beam of mode B propagate through $\sim 40$~mm of N-BK7, with a transmission of 92\% per 10~mm at 810~nm, leading to a total transmission through the prism of $71$\%. The prism is also uncoated and at the external angle we use of $\sim 20^\circ$ to the normal of the prism surface there is $3-5$\% loss on input to the prism (using Fresnel's equations and depending on the polarisation set) and similarly for the output from the prism. This leads to a combined transmission of $64$\% during total internal reflection, in contrast to $T=100$\% expected ideally. As the external angle increases in order to move along the SPP resonance dip and closer to the inflection point, the transmission drops further slightly due to increased reflections at the input/output prism surfaces. The remaining decrease of the transmission to 10\% can be attributed to the detector efficiency and the efficiency of coupling collected light into the fibers before detection. This final decrease was confirmed in a control experiment where the prism was not present, giving an overall transmission $\sim 20$\%. Thus, we have $T=0.2 \times 0.64=13$\%. The remaining additional few percent may be attributed to the surface roughness of the gold, which has a small influence~\cite{Kanso07}, and other surface reflections, such as those at the interface of the index matching oil.

The incident angle of photons entering the plasmonic sensor is increased away from the total internal reflection point toward the dip center such that the transmission of the sensor is reduced by roughly 40\% and we are operating close to the point on the dip that is most sensitive to refractive index changes. Thus, the initial transmission (with deionized water as the buffer solution) is $0.1 \times (1-0.4)=6$\% and rises as BSA is added and the binding reaches a steady state, pushing the sensor back toward total internal reflection where the transmission is about 10\%. The dashed line in Fig.~\ref{fig1}~(d) is a nonlinear fit to the points, the details of which are given in the next section. As 1.5\% BSA was enough to increase the transmission back to its maximum, we did not consider higher concentrations, as these would have led to transmission values that are not consistent with a linear response to the refractive index change (see right side of inset in Fig.~\ref{fig1}~(a)).
\begin{figure*}[t]
\centering
\includegraphics[width=17cm]{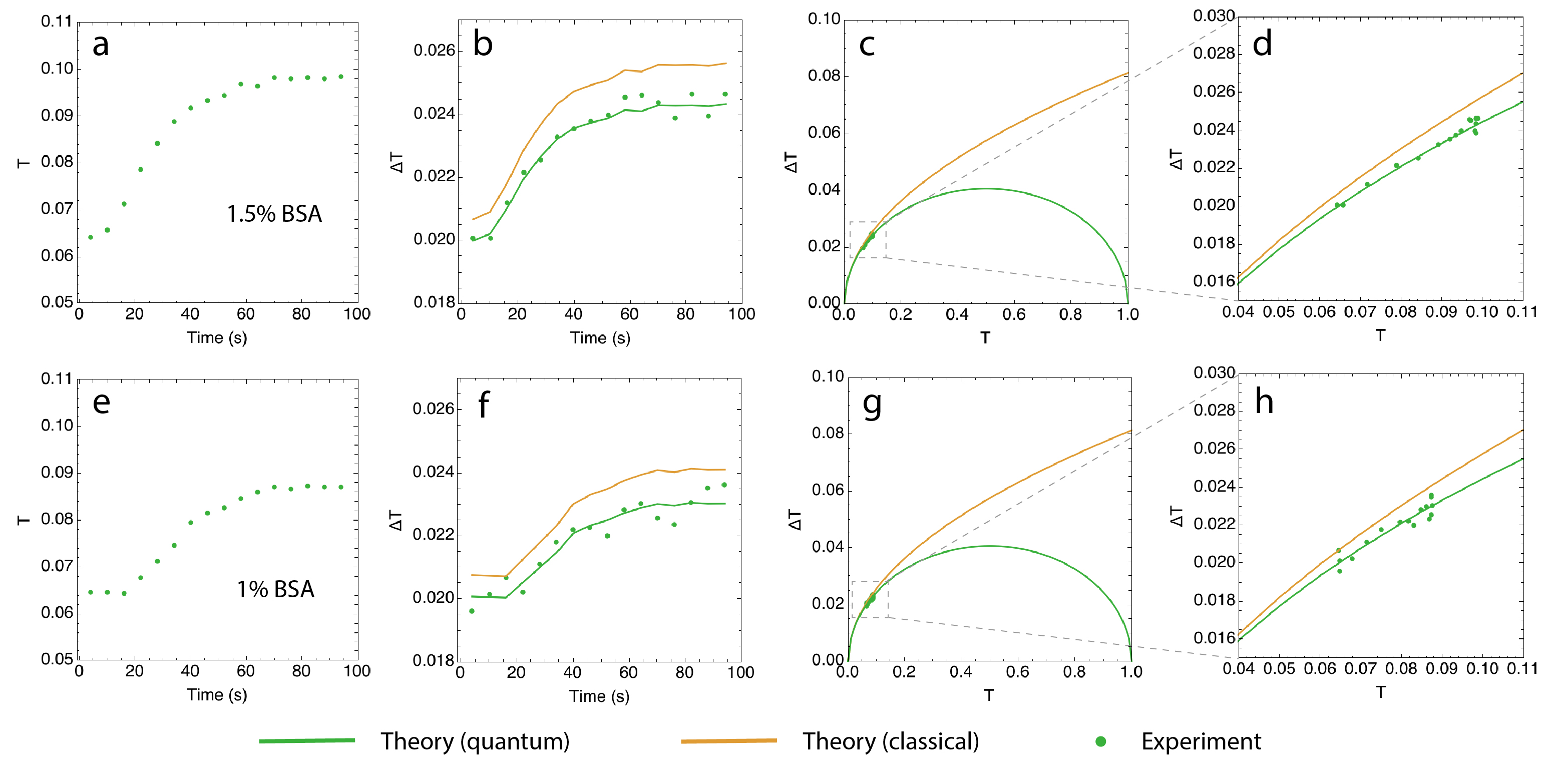}
\caption{Sensorgram and its precision for an injection of BSA with a 1.5\% and 1\% concentration above the gold surface. Top row shows 1.5\% BSA and bottom row shows 1\% BSA. (a) Measured sensorgram $\langle T \rangle$ for 1.5\% BSA. (b) Standard deviation $\Delta T$ associated with $\langle T \rangle$ for each point in time. The green (orange) line is the expected quantum (classical) standard deviation $\Delta T_{\rm quantum}$ ($\Delta T_{\rm classical}$).  (c) Values of $\Delta T$ that correspond to the values of $\langle T \rangle$ with the temporal aspect of the sensorgram removed. (d) Zoomed in region of panel (c) highlighting the gap between the classical and quantum cases. (e) Measured sensorgram $\langle T \rangle$ for 1\% BSA. (f) Standard deviation $\Delta T$ associated with $\langle T \rangle$ for each point in time. (g) Values of $\Delta T$ that correspond to the values of $\langle T \rangle$ with the temporal aspect of the sensorgram removed. (h) Zoomed in region of panel (g) highlighting the gap between the classical and quantum cases.}
\label{fig2} 
\end{figure*}

The experimental precision obtained from Eq.~\eqref{eq:expDT} can be compared with the theoretical model for the ideal case of single photons (quantum) and coherent states with mean photon number of one (classical), which are given by~\cite{Lee2018}
\be
\label{eqn:quant}
 \Delta T_{\rm quantum} = \sqrt{\frac{\langle T \rangle(1-\langle T \rangle)}{\nu}}.
\ee
and
\be
\label{eqn:clas}
\Delta T_{\rm classsical} = \sqrt{\frac{\langle T \rangle}{\nu}}.
\ee
We compare our experimental precision with these theoretical predictions in the next section. It is important to note that the above equations have a $N^{-1/2}$ dependence in general~\cite{Lee2018}, where $N$ is the mean photon number of the $N$-photon number state (quantum) and coherent state (classical).

%%%%%%%%%%%%%%%%%%%%%%%%%%%%
%%%%%%%%%%%%%%%%%%%%%%%%%%%%
%%%%%%%%%%%%%%%%%%%%%%%%%%%%
%%%%%%%%%%%%%%%%%%%%%%%%%%%%
\section{Results}

%%%%%%%%%%%%%%%%%%%%%%%%%%%%
%%%%%%%%%%%%%%%%%%%%%%%%%%%%
\subsection{Transmission measurement}

Interaction kinetics can be divided into three main stages: association, steady state and dissociation~\cite{Xiao19}. The association stage mainly involves the binding of ligands to receptors to form receptor-ligand complexes, although some unbinding (release of ligands) also occurs. In the present context, a BSA molecule plays the role of a ligand and an area on the gold surface is the receptor. The steady state stage is where equilibrium is reached with the number of ligands binding equaling the number that are unbinding. The dissociation stage involves the irreversible unbinding of ligands and receptors, which occurs when the BSA solution is replaced by a buffer solution in a process called elution. In this work we focus on the association and steady state stages, as these are sufficient to extract out all the kinetic parameters for the interaction, including the dissociation parameter~\cite{Xiao19}.
\begin{figure*}[t]
\centering
\includegraphics[width=17cm]{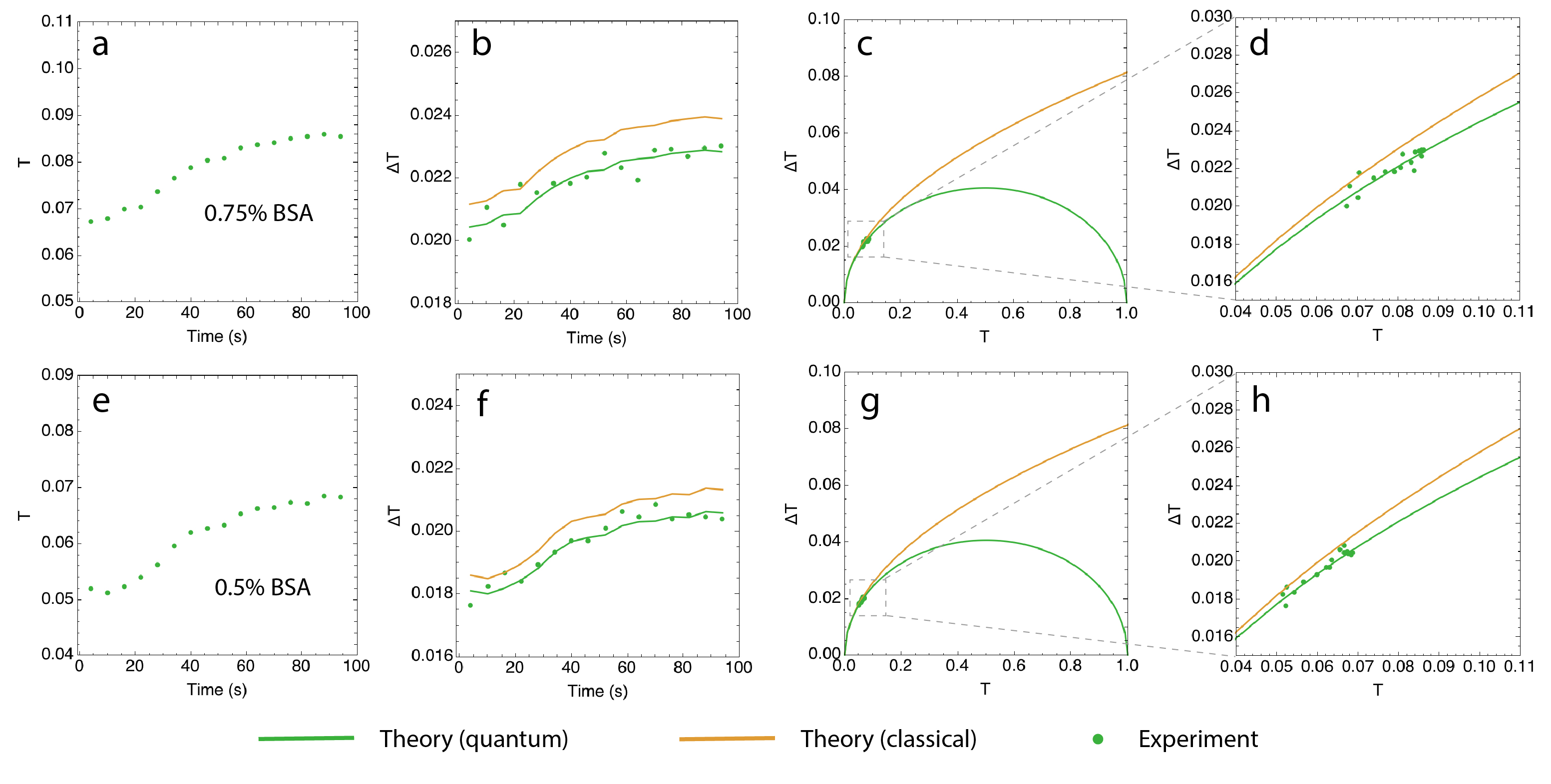}
\caption{Sensorgram and its precision for an injection of BSA with a 0.75\% and 0.5\% concentration above the gold surface. Top row shows 0.75\% BSA and bottom row shows 0.5\% BSA. (a) Measured sensorgram $\langle T \rangle$ for 0.75\% BSA. (b) Standard deviation $\Delta T$ associated with $\langle T \rangle$ for each point in time. (c) Values of $\Delta T$ that correspond to the values of $\langle T \rangle$ with the temporal aspect of the sensorgram removed. (d) Zoomed in region of panel (c) highlighting the gap between the classical and quantum cases. (e) Measured sensorgram $\langle T \rangle$ for 0.5\% BSA. (f) Standard deviation $\Delta T$ associated with $\langle T \rangle$ for each point in time. (g) Values of $\Delta T$ that correspond to the values of $\langle T \rangle$ with the temporal aspect of the sensorgram removed. (h) Zoomed in region of panel (g) highlighting the gap between the classical and quantum cases.}
\label{fig3} 
\end{figure*}

When BSA is added to the region above the gold surface the value of the permittivity $\epsilon_a$ increases in that region due to binding of BSA molecules to the gold. The transmission $T$ increases and gains a time dependence, as can be seen in Fig.~\ref{fig1}~(b). During this association stage, BSA molecules bind and unbind with the gold surface and the steady state stage is eventually reached. 

The concentration of the receptor-ligand complex $[C]$ and the transmission of the sensor $T$ can be linked by the refractive index, $n_a=\sqrt{\epsilon_a}$, of the region above the gold surface. For a fixed incidence angle, an increase in the complex concentration $[C]$ increases the refractive index and thus $T$, as shown in the inset of Fig.~\ref{fig1}~(a). A mathematical model for the association and steady state stages of the interaction is given by~\cite{Xiao19} 
\be
T(t) ={T}_{\infty}(1-{e}^{-{k_s t}}),
\label{sensorgrameqm}
\ee
where $T_\infty$ is a constant determined by the initial concentration of the ligands, $[{L_0}]$, and receptors, $[{R_0}]$, the thickness of the ligand layer above the gold surface, $\ell$, and the affinity of the receptor-ligand interaction $K_A=\frac{{k_a}}{{k_d}}$. Here, ${k_a}$ is the association parameter in ${\rm M}^{-1} {\rm s^{-1}}$ (per molarity per second) and ${k_d}$ is the dissociation parameter in ${\rm s^{-1}}$. In Eq.~\eqref{sensorgrameqm}, the parameter ${k_s}={k_a}[{L_0}]+ {k_d}$ represents the `observable rate' in units of ${\rm s^{-1}}$. From the measured sensorgram of $T$ in the experiment a nonlinear fit to the model given in Eq.~\eqref{sensorgrameqm} is performed using a Gauss-Newton method in order to extract out $k_s$. We start by focusing on the estimation of the observable rate parameter, $k_s$, before considering the estimation of the association and dissociation parameters. 

In Fig.~\ref{fig2}~(a) we show the transmission $T$ measured for an injection of BSA with a 1.5\% concentration above the gold surface. Each point is the average $\langle T \rangle$ from $\mu=2 \times 10^3$ sets of measurements within 6 seconds, each set of measurements having $\nu=150$ probes, as described in the previous section. In Fig.~\ref{fig2}~(b) we show the standard deviation $\Delta T$ for the mean of the $\mu$ sets for each point in time. The solid green line gives the expected theory values for using single-photons based on substituting $\langle T \rangle$ from Fig.~\ref{fig2}~(a) into Eq.~\eqref{eqn:quant}. The solid orange line gives the expected theory values for a coherent state which are obtained by substituting $\langle T \rangle$ from Fig.~\ref{fig2}~(a) into Eq.~\eqref{eqn:clas}. The experimental points are clearly in line with the expected single-photon case, demonstrating a smaller standard deviation $\Delta T$ and therefore an enhancement in the estimation precision. The close match between the experiment and single-photon theory prediction confirms that the noise in the experiment is mainly due to the statistics of the single-photons and that technical noise at low frequencies (slowly varying) in the sensor, such as laser and vibrational fluctuations are a small contribution to the observed precision.

It should be noted that in the classical case of a coherent state, the precision is set by the shot noise, which in general (mean photon number $>1$) is inversely proportional to the square root of the intensity~\cite{Lee2021}. Thus, in principle, one could simply increase the intensity per probe state, {\it i.e.}, mean photon number, (or alternatively the rate of probing at fixed intensity per probe state) and thereby decrease the noise to obtain a better precision. The important point here is that for a fixed mean photon number per state (in this case one) and fixed rate of probing, the quantum case always outperforms the classical case in terms of giving a smaller precision. Consequently, one can obtain the same estimation precision as the classical case using a quantum state with a reduced intensity. This is important when the biological sample is photosensitive~\cite{Casacio2021}, or the sensor is operating close to its intensity limit in terms of linear response~\cite{Homola2009,Kaya2013}. It is therefore this setting where our quantum plasmonic sensor would provide a practical advantage.

To put the relation between $\langle T \rangle$ and $\Delta T$ in context and highlight the quantum advantage, in Fig.~\ref{fig2}~(c) we plot the values of $\Delta T$ that correspond to the values of $\langle T \rangle$, thereby removing the temporal aspect of the sensorgram. As before, the solid green (orange) line gives the expected theory value for the single-photon (coherent state) case. The difference in $\Delta T$ between the quantum and classical case is small for low values of $\langle T \rangle$, and so in Fig.~\ref{fig2}~(d) we show a zoomed in plot, highlighting the reduction more clearly with all experimental points closest to the expected quantum case. It is important to note that a higher overall transmission in our setup would widen the gap between the classical and quantum case, and provide an even better enhancement in the precision, but already with the current setup one can see that this gap is present. There is a potential for further enhancement by improving the detector efficiency, reducing the prism size, adding an anti-reflection coating on the prism and optimizing the coupling of the light into the fibers before detection.

In Fig.~\ref{fig2}~(e)-(h) we show the results for an injection of BSA with a 1\% concentration above the gold surface. As in the previous case, the experimental points are in line with the expected single-photon case, demonstrating an enhancement in the estimation precision. In Fig.~\ref{fig3}~(a)-(d) ((e)-(h)) we show the results for an injection of BSA with a 0.75\% (0.5\%) concentration above the gold surface. The experimental points are roughly in line with the expected single-photon case, although some are close to the classical coherent state case. The reason for this is because a lower concentration of BSA results in a smaller deviation of $T$ in the sensorgram, which corresponds to a region where there is a smaller gap between the expected classical and quantum standard deviations. This can be seen by comparing Fig.~\ref{fig2}~(d) and (h), and Fig.~\ref{fig3}~(d) and (h). Despite the small difference between the precisions an overall enhancement in the estimation precision can be seen.

To highlight better the improvement in the precision, in Fig.~\ref{fig4}~(a) to (d) we show the enhancement of the precision, $\Delta T_{\rm classical}/\Delta T$, for the different concentrations over time as the sensorgram changes. These plots are obtained using the values of $\Delta T$ shown in Fig.~\ref{fig2}~(b) and (f) for 1.5\% and 1\%, and from the values in Fig.~\ref{fig3}~(b) and (f) for 0.75\% and 0.5\%. The shot noise limit (SNL) is set by the classical case and represents a benchmark, above which we can say there is a `quantum enhancement'. The dashed green line is the expected theory value of the enhancement based on the value of $\langle T \rangle$, and using Eqs.~(\ref{eqn:quant}) and (\ref{eqn:clas}).

%%%%%%%%%%%%%%%%%%%%%%%%%%%%
%%%%%%%%%%%%%%%%%%%%%%%%%%%%
\subsection{Estimation of kinetic parameters}

We now consider how the enhancement in the estimation precision of the transmission $T$ translates to an enhancement in the precision of measuring kinetic parameters, as predicted in a recent theoretical work~\cite{Mpofu2021}. For a given concentration, in order to extract out the kinetic parameter $k_s$ from the sensorgram $T(t)$, {\it e.g.}, Fig.~\ref{fig2}~(a) for 1.5 \% BSA, we must take into account that the sensorgram has noise, $\Delta T$, associated with the value of $T(t)$ at each point, {\it e.g.}, Fig.~\ref{fig2}~(b). Thus a simple fit to the mean $T(t)$ sensorgram and subsequent extraction of the kinetic parameter will not provide information about the estimation precision of that parameter. We therefore perform a bootstrap sampling of our data as follows. For each point in time we set the value of $T(t)$ to be $T_i$, with $i$ randomly chosen from $\mu$ sets, {\it i.e.}, $i=1, \cdots, \mu$. Here, the transmission $T_i$ is measured in our experiment from a set of $\nu$ measurements at that point in time. This produces a single noisy sensorgram from our data. We repeat this sensorgram generation $m=175$ times, giving a set of 175 noisy sensorgrams. The signal-to-noise ratio in each of these sensorgrams is unfortunately too small to allow a fit of the model in Eq.~\eqref{sensorgrameqm} due to the low value of $\nu$ -- see Fig.~\ref{fig1}~(d) for a representation of how the signal values ($T_i$) at each point in time of a single sensorgram vary due to the large noise ($\Delta T$) given by the error bar. The 175 sensorgrams are therefore averaged into a single mean sensorgram. We then apply a nonlinear fit of the model given in Eq.~\eqref{sensorgrameqm} to the mean sensorgram in order to extract out a single $k_s$ value. The fit shown in Fig.~\ref{fig1}~(d) is an example of one of these nonlinear fits. For the nonlinear fit we use Mathematica's {\sf NonLinearModelFit} function with the Levenberg-Marquardt method, which interpolates between the Gauss-Newton and gradient descent method. This method is more robust than the Gauss-Newton method on its own and represents a refined Gauss-Newton method using a trust region approach~\cite{Nocedal2006}.

\begin{figure}[t]
\centering
\includegraphics[width=8cm]{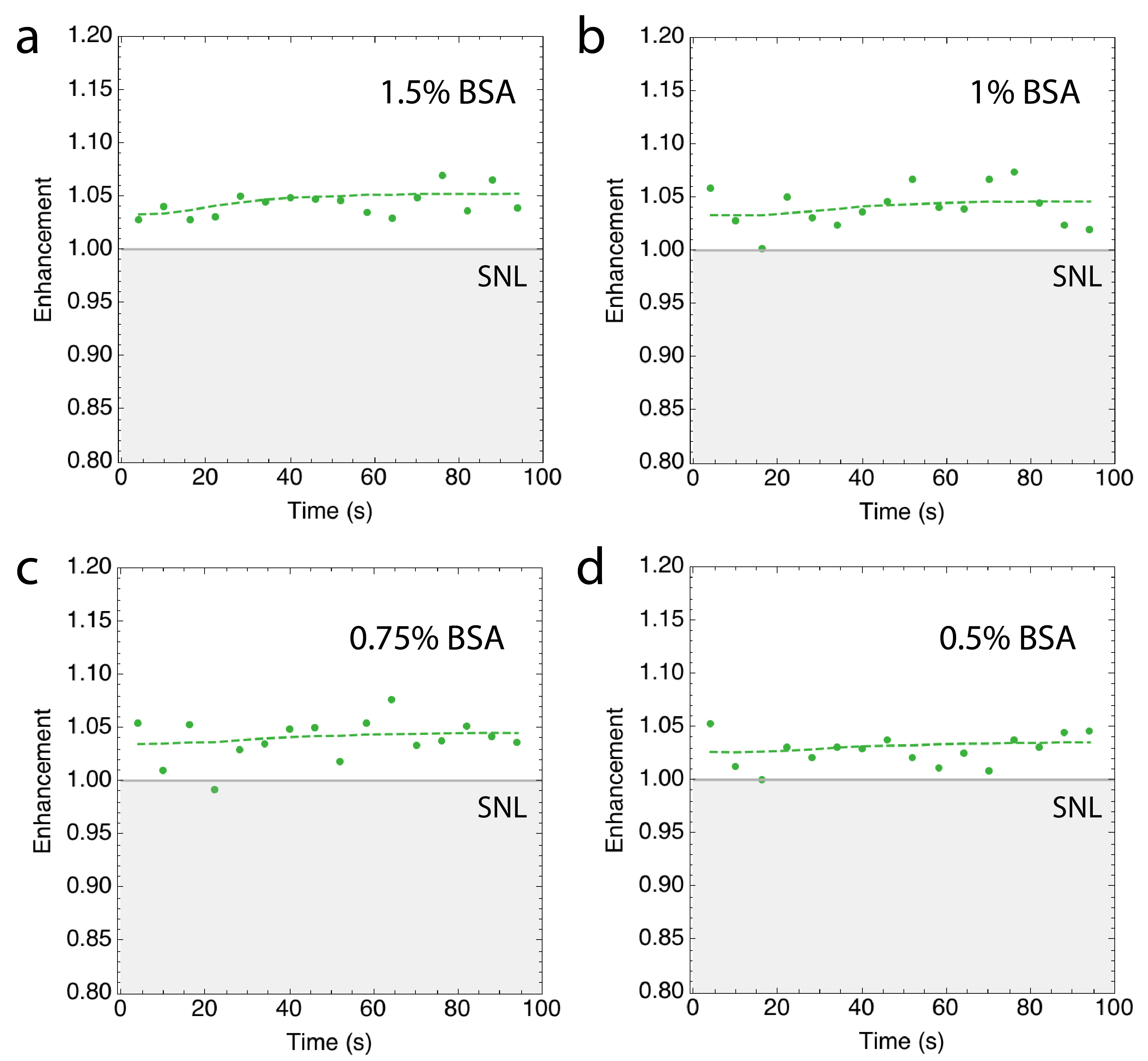}
\caption{Enhancement of the precision, $\Delta T_{\rm classical}/\Delta T$, for the different concentrations of BSA. (a) 1.5\% BSA. (b) 1\% BSA. (c) 0.75\% BSA. (d) 0.5\% BSA. The shot noise limit (SNL) is equal to 1 and set by the classical case. The dashed green line is the expected theory value of the enhancement based on the value of $\langle T \rangle$.}
\label{fig4} 
\end{figure}

In order to quantify the estimation precision from our statistical data processing, we repeat the above sampling process $p=15 \times 10^{3}$ times to get $p$ values of $k_s$. We then calculate the mean $\bar{k}_s$ and standard deviation, $\Delta \bar{k}_s$. This gives the estimate and the precision in the estimation of $k_s$, respectively, for a {\it single set} of $m=175$ noisy sensorgrams from our data. The value of $p$ was set by gradually increasing it to $15 \times 10^{3}$, where it was found to give a stable mean and standard deviation for $k_s$. For the different concentrations the value of the estimate $\bar{k}_s$ and its precision $\Delta \bar{k}_s$ are given in Tab.~\ref{tab1} for the case of single photons from our experiment and the classical expected case. In the classical case, we followed the same bootstrap procedure as detailed above, but obtained $T_i$ by taking the value of $\langle T \rangle$ obtained from the experiment at a given point in time and adding Gaussian noise to it with a standard deviation of $\sqrt{\langle T \rangle}$, in line with Eq.~\eqref{eqn:clas}. The table clearly shows that the single-photon case gives a better precision for all concentrations. However, the gap between the precision decreases as the concentration decreases, which is a result of the enhancement in the precision of $T$ decreasing, as seen in Fig.~\ref{fig4}. The largest enhancement in the precision of estimating $k_s$ is $0.42/0.39=1.077$ for 1.5\% BSA, corresponding to a percentage change, or improvement, in the precision of $(0.42-0.39)/0.42=7.1\%$ using single photons.

The mean values of $\bar{k}_s$ are slightly higher for the classical case compared to the quantum case for all concentrations, which is a small bias effect predicted from theory when there is a corresponding larger precision $\Delta \bar{k}_s$~\cite{Mpofu2021}. Despite this, the mean values of the quantum and classical cases are consistent with each other within their precision bounds.

We now turn our attention to the extraction of the association parameter, $k_a$, and the dissociation parameter, $k_d$. First, in order to obtain the dissociation parameter, usually an elution process is used, where the BSA solution is replaced by a buffer solution and irreversible unbinding of ligands and receptors occurs at the rate $k_d$. Then, with a knowledge of $k_d$, $k_s$ and $[{L_0}]$, the association parameter can be obtained from the relation ${k_s}={k_a}[{L_0}]+ {k_d}$. However, in many binding interactions the elution process is not ideal due to various factors, including diffusion~\cite{Karlsson1999} and side-hindrance~\cite{Brewer2005}. An alternative method that can be used is the so-called double reciprocal method based on the following equation~\cite{Xiao19}
\be
\frac{1}{T_\infty}=\alpha K_A^{-1} \frac{1}{[{L_0}]}+\alpha,
\label{dr}
\ee
where $\alpha$ is the $y$-intercept of a plot of $1/T_\infty$ vs $1/[{L_0}]$. From the $y$-intercept and gradient of a double reciprocal plot we obtain $K_A$, and using this we can write ${k_s}=K_A k_d[{L_0}]+ {k_d}$. With a knowledge of $k_s$, $[{L_0}]$ and $K_A$ we then obtain the dissociation parameter $k_d$. Finally, from the relation ${k_s}={k_a}[{L_0}]+ {k_d}$ we obtain the association parameter $k_a$.
\begin{table}[b]
\begin{tabular}{ccc}
\hline
\hline
\phantom{p} & \qquad \qquad ~~~~Quantum & Classical \\
\phantom{p} &
\begin{tabular}{ccccc}
\phantom{p} & 1.5\% & 1\% & 0.75\% & 0.5\% \\
$\bar{k}_s$ ($10^{-2}$) & 4.11 & 4.00 & 2.97 & 1.81 \\
$\Delta \bar{k}_s$ ($10^{-2}$) & 0.39 & 0.61 & 0.71 & 0.49 \\
\end{tabular}
&
\begin{tabular}{ccccc}
 1.5\% & 1\% & 0.75\% & 0.5\% \\
 4.40 & 4.12 & 2.86 & 2.00 \\
 0.42 & 0.65 & 0.73 & 0.50 \\
\end{tabular}
\\
\hline
\hline
\end{tabular}
\caption{The value of the estimate $\bar{k}_s$ in units of ${\rm s^{-1}}$ and its precision $\Delta \bar{k}_s$ at the different concentrations for the case of single photons from our experiment (quantum) and a coherent state (classical).}
\label{tab1}
\end{table}

In order to use the double reciprocal method, we must calculate the actual concentration of BSA, $[{L_0}]$, in the cavity region above the gold surface for each concentration injected. The cavity region is initially filled with 0.5~ml of deionized water and for each of the respective sensorgrams, 0.13~ml of BSA is injected with the concentrations stated previously: 1.5\%, 1\%, 0.75\% and 0.5\%. Taking the first concentration as an example (1.5\%), we prepare the BSA before injection by dissolving 0.15~g of dry powder BSA into 10~ml of deionized water (1 hour before use). This gives the initial concentration $[L_i]=0.15/(66430\times 10\times 10^{-3})=2.258\times 10^{-4}$~mol/l (moles per liter, or molarity ${\rm M}$), where the molar mass of BSA is taken to be 66430~g/mol~\cite{Hirayama1990}. The concentration in the cavity after injection is then $[L_0]=[L_i]V_i/V_0=2.258\times 10^{-4}(0.13/(0.13+0.5))=4.659\times10^{-5}$~mol/l. A similar calculation can be done for the other concentrations. In summary, we have $[L_0]=4.659\times10^{-5}$~mol/l (1.5\%), $3.106\times10^{-5}$~mol/l (1\%), $2.330\times10^{-5}$~mol/l (0.75\%) and $1.553\times10^{-5}$~mol/l (0.5\%). 

\begin{figure}[t]
\centering
\includegraphics[width=8.7cm]{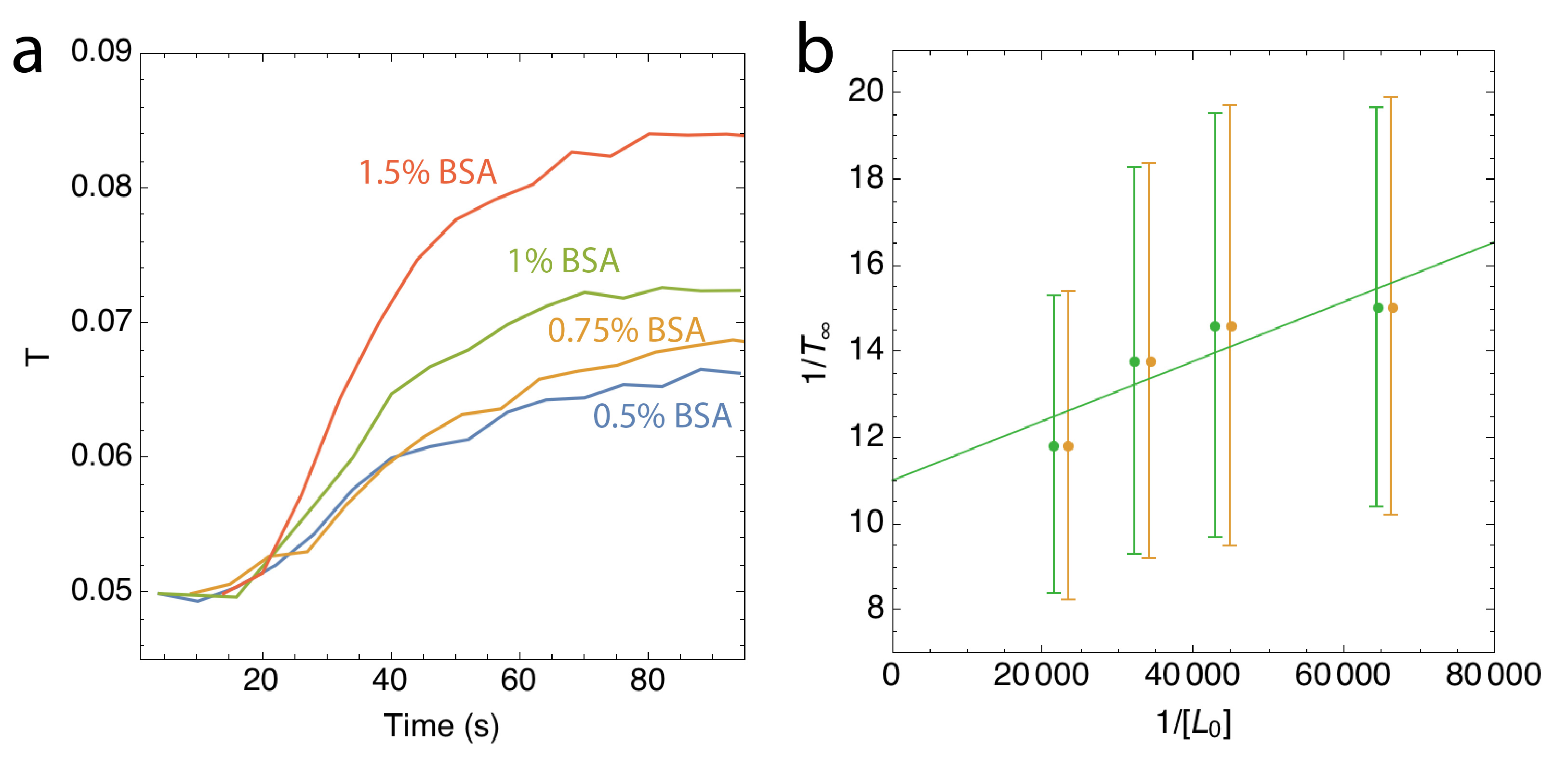}
\caption{Estimation of kinetic parameters $k_a$ and $k_d$. (a) Adjusted sensorgrams for the different BSA concentrations injected into the region above the gold surface. (b) Double reciprocal plot used to extract out the affinity $K_A$ for the interaction (see text for details), leading to estimates for the association ($k_a$) and dissociation ($k_d$) parameters. The values of $T_\infty$ in panel (b) are taken from the values for $T$ in panel (a) at 94 seconds. The error bars for the classical case have been shifted slightly from those for the quantum case in order to compare them more easily. The solid green line is a linear fit to the points.}
\label{fig5} 
\end{figure}

The values of $T_\infty$ corresponding to the above $[L_0]$ values are taken from an adjusted sensorgram plot, where each sensorgram is shifted in time so that the injection points of all sensorgrams match up. This is done as the injection of BSA is performed manually and the time at which it occurs after the time tagging collection is initiated is challenging to keep constant. The transmissions are also shifted slightly so that the initial transmissions $T(0)$ all match up. This shift is done as it was not possible in the experiment to return to the exact same position on the inflection curve for each concentration due to temperature and alignment fluctuations. The adjusted sensorgrams are shown in Fig.~\ref{fig5}~(a), which are simply adjusted versions of Fig.~\ref{fig2}~(a) and (e), and Fig.~\ref{fig3}~(a) and (e). The mean value of $T_\infty$ is taken from the value at 94 seconds, corresponding to the mid-point of the 6 second period from 91-97 seconds for the different $[L_0]$ sensorgrams, along with with its standard deviation $\Delta T_\infty$. The resulting double reciprocal plot is shown in Fig.~\ref{fig5}~(b) together with the expected classical case obtained by using Eq.~\eqref{eqn:clas}, as previously done for the parameter $k_s$. The mean value of $T_\infty$ is the same in both the quantum and classical case, however the standard deviations in the quantum case are slightly smaller. The signal-to-noise ratio in the double reciprocal plot for either the quantum or classical case is too small to allow a fit of a linear model to find $k_d$ and $k_a$ (see Eq. \eqref{dr}) due to the low value of $\nu$ giving large error bars. This is a similar problem to before when we attempted to apply a nonlinear fit to a noisy sensorgram to find $k_s$.

To extract out estimates of $k_d$ and $k_a$ along with their estimation precision we perform a bootstrap sampling of the experimental data, as before. We do this by taking a value of $T_\infty$ from a set of $\nu$ measurements from our total of $\mu$ sets for each concentration, with $[L_0]$ taken to be exact as the error is negligible in relative terms ($\simeq 10^{-7}$)~\cite{L0error}. We then repeat the sampling process $m=175$ times for each concentration in order to be consistent with the $k_s$ estimation procedure already performed. From the mean values of $T_\infty$ we make a double reciprocal plot and find $K_A$ from a fit to the equation $y=mx+c$ using the Mathematica function {\sf LinearModelFit}. We repeat this process $p=15 \times 10^3$ times, again to be consistent with the case of $k_s$. We also take $p$ values of $k_s$ from its distribution for the concentration $[L_0]=1.5\%$, as this gives the largest enhancement in the precision of $\bar{k}_s$. The $k_s$ values together with the $K_A$ values give $p$ values of $k_d$ and $k_a$ using the extraction method described above. We then calculate the means $\bar{k}_d$ and $\bar{k}_a$, and standard deviations $\Delta \bar{k}_d$ and $\Delta \bar{k}_a$. 

For the dissociation parameter, we find $\bar{k}_d=0.0100~{\rm s}^{-1}$ and $\Delta \bar{k}_d=0.0015~{\rm s}^{-1}$ for single photons and $\bar{k}_d=0.0110~{\rm s}^{-1}$ and $\Delta \bar{k}_d=0.0022~{\rm s}^{-1}$ for the classical case. For the association parameter we find $\bar{k}_a=672.2~{\rm M}^{-1} {\rm s^{-1}}$ and $\Delta \bar{k}_a=68.8~{\rm M}^{-1} {\rm s^{-1}}$ for single photons and $\bar{k}_a=708.1~{\rm M}^{-1} {\rm s^{-1}}$ and $\Delta \bar{k}_a=78.5~{\rm M}^{-1} {\rm s^{-1}}$ for the classical case. The enhancement in the estimation precision of $k_a$ is 1.14, which corresponds to an improvement in the precision of $12.4 \%$, while the enhancement in the estimation precision of $k_d$ is the highest at 1.47, which corresponds to an improvement in the precision of $31.8 \%$.

As a check of our results with those found in other studies using classical light, the value of the affinity for the interaction between BSA and a flat gold surface can be found in the literature to be $K_A=k_a/k_d=0.02~\mu$M~\cite{Brewer2005}. The estimate we have obtained is $\bar{K}_A=0.069~\mu$M with a precision of $\Delta \bar{K}_A=0.011~\mu$M. However, the value obtained in the work from Brewer {\it et al.}~\cite{Brewer2005} is not a reliable estimate. Indeed, in another work, Boulos et al.~\cite{Boulos2013}, it is mentioned that one can find a wide range of values in the literature for the affinity of BSA interacting with gold, spanning several orders of magnitude. Thus, although it is not clear to what extent our values of $k_d$ and $k_a$ are accurate to the true values, we have shown that by using quantum states the precision in the estimates of the kinetic parameters is improved by up to $31.8$\% compared to the classical case. 

%%%%%%%%%%%%%%%%%%%%%%%%%%%%
%%%%%%%%%%%%%%%%%%%%%%%%%%%%
%%%%%%%%%%%%%%%%%%%%%%%%%%%%
%%%%%%%%%%%%%%%%%%%%%%%%%%%%
\section{Summary}

We have reported a proof-of-principle experiment that demonstrates a quantum enhancement in the precision of estimating kinetic parameters. We used single photons as the quantum light source, which were sent into a plasmonic resonance sensor set up to monitor the interaction of the protein BSA to gold. As BSA is a protein that is capable of binding to many types of antibodies and drugs, this is an informative first test case in the practical study of whether a quantum enhancement can be achieved in the precision of measuring kinetic parameters. Due to the reduced noise of the single-photon statistics we found that an improvement in the precision of up to 31.8\% in the values of kinetic parameters is possible, confirming recent theoretical predictions~\cite{Mpofu2021}. This work shows that quantum light sources can realistically be used for sensing of kinetic parameters with an improved precision compared to a classical approach. Our results may open up new possibilities for designing quantum-based sensors for biochemical research.

Several improvements to our setup would enable a larger enhancement in the precision to be obtained. The key to the improvement is increasing the overall transmission in the setup~\cite{Lee2018}. This can be achieved by increasing the detector efficiency (currently at ${\sim}60\%$), increasing the transmission through the prism when off resonance (currently at ${\sim}64\%$) by decreasing the prism size and adding anti-reflection coatings, using a source of pairs of photons with an improved coincidence-to-singles ratio~\cite{Lee2018} and optimised coupling into the collection fibers before detection. With these improvements, the enhancement in precision may be pushed much higher~\cite{Lee2018,Peng2020,Zhao2020}. Another direction to improve the precision would be to use alternative quantum states~\cite{Lee2021}, such as the two-mode squeezed vacuum state and two-mode squeezed displaced state, which offer a similar enhancement, but due to a potential increase in intensity per state (mean photon number), they would improve the overall precision for the same rate of probing~\cite{Mpofu2021}. On the other hand, the rate of probing could be increased in our setup using a brighter source of single photons~\cite{Eisaman2010}. In the experiment we have used a low value of $\nu=150$, but a brighter source would allow $\nu$ to be increased and lead to an increase in the overall intensity. This would also reduce the integration time of measurements and thereby suppress technical noise at low frequencies in the sensor, such as laser and vibrational fluctuations, which are additional smaller sources of noise added to the shot noise and contribute to the observed precision. One could then potentially study the $\nu$ dependence of the estimation precision~\cite{Mpofu2021}.

%%%%%%%%%%%%%%%%%%%%%%%%%%%%
%%%%%%%%%%%%%%%%%%%%%%%%%%%%
%%%%%%%%%%%%%%%%%%%%%%%%%%%%
%%%%%%%%%%%%%%%%%%%%%%%%%%%%
\begin{acknowledgments}
This research was supported by the South African National Research Foundation, the National Laser Centre and the South African Research Chair Initiative of the Department of Science and Innovation and National Research Foundation. C. L. is supported by a KIAS Individual Grant (QP081101) via the Quantum Universe Center at Korea Institute for Advanced Study and Korea Research Institute of Standards and Science (KRISS–GP2022-0012). G. E. M. M. and H. G. K. thank UKZN for financial and administrative support.
\end{acknowledgments}

%\appendix
%\section{Appendixes}

\nocite{*}
%\bibliography{aipsamp}% Produces the bibliography via BibTeX.

\begin{thebibliography}{99}
\bibitem{L0error} The equation for the concentration is $[L_0]=[L_i]V_i/V_0$. The error in the measurement of $V_i$ is $\Delta V_i=10^{-5}$~l and the error in $V_0$ is $\Delta V_0=2\times10^{-5}$~l (sum of two volumes). The initial concentration $[L_i]=M_{BSA}/(V_{BSA}M_{BSAm})$, where $M_{BSA}$ is the mass of BSA dry powder, $V_{BSA}$ is the volume of deionized water used and $M_{BSAm}$ is the molar mass of BSA. The errors are $\Delta M_{BSA}=10^{-4}$, $\Delta V_{BSA}=10^{-5}$ and $\Delta M_{BSAm}\simeq1$, and we have using error propagation (taking 1.5\% as an example) an error of $\Delta [L_0]\simeq 10^{-7}$, which is $\simeq0.2\%$ of the value of $[L_0]$.
\end{thebibliography}

\end{document}